\documentclass[twocolumn, prl, amsmath,amssymb]{revtex4-2}
\usepackage[pdftex]{graphicx}
\usepackage{dcolumn}
\usepackage{bm}
\usepackage[T1]{fontenc}
\usepackage[utf8]{inputenc}
\usepackage{natbib}
\usepackage{hyperref}
\usepackage{xcolor}
\usepackage{url}
\usepackage{ulem}
\usepackage{float}

{\rule{20pt}{1ex}\endlist}
\let\oldmarginpar\marginpar
\renewcommand\marginpar[1]{\-\oldmarginpar[\raggedleft\footnotesize #1]%
{\raggedright\footnotesize #1}}

\begin{document}


\title{Guided mechanical self-assembly of bubbles in fiber arrays}
\author{Marwan Chammouma$^{1}$} \thanks{These two authors contributed equally}

\author{Manon Jouanlanne$^{1}$} \thanks{These two authors contributed equally}

\author{Antoine Egelé$^{1}$}
\author{Damien Favier$^{1}$}
\author{Jean Farago$^{1}$}
\author{Aur\'elie Hourlier-Fargette$^{1}$}
\affiliation{
$^1$ Université de Strasbourg, CNRS, Institut Charles Sadron UPR22, F-67000 Strasbourg, France\\}

\date{\today}

\begin{abstract}

Spontaneous mechanical self-assembly of monodisperse bubbles generally leads to disordered foams at low density: producing crystalline structures such as Kelvin foams has proven to be challenging experimentally, despite them being a minimum of energy. Here we show how bubbling in different fiber arrays controls foam architectures through a guided mechanical self-assembly. Analyzing X-ray tomography scans of solidified polymer foams using Steinhardt’s parameters highlights clear signatures of Kelvin and hexagonal close packing crystalline foams, opening a novel route towards ordered hierarchical materials.

\end{abstract}

\maketitle


 Properties of liquid and solid foams are highly dependent on their geometrical features \cite{Drenckhan2015, Andrieux2018}, attracting interest into strategies to control the way bubbles spontaneously assemble in three dimensions. Packing bubbles of equal sizes into a foam generally leads to disordered structures, unless specific care is given to the obtention of crystalline architectures \cite{Drenckhan2010}. Interfacial energy minimization - leading to Plateau's laws - guide the assembly \cite{Plateau1873}, but experimentally, structures often correspond to local minima of energy, rearrangements of the bubbles being too costly in energy to occur. As a striking example, Kelvin predicted in 1887 that packing truncated octahedrons made of 6 square faces and 8 hexagons to pave the space in 3D would minimize the surface and thus the interfacial energy \cite{Thomson1887}, but such cells were elusive experimentally. In the historical work of Matzke in the 40s \cite{Matzke1946}, observing under a microscope numerous cells in a foam prepared bubble by bubble, no Kelvin cell had been observed. One isolated instance of observation occured ten years later \cite{Dodd1955}, and Kelvin cells assemblies have been observed more recently as detailed below at the cost of complicated strategies \cite{Rosa2002, Hohler2008, vanderNet2007, Meagher2013, Saadatfar2008, Skrypnik2023}. Weaire and Phelan predicted theoretically another structure composed of an arrangement of two types of bubble shapes \cite{Weaire1994}, that provides a more efficient partition of space than the Kelvin cell, which was even harder to obtain experimentally, and was achieved only a decade ago \cite{Gabbrielli2012}. Those two examples of experimental hunts for Kelvin and Weaire-Phelan foams illustrate the difficulty to experimentally obtain hierarchical ordered foams with desired structures.

Two main general strategies to reach Kelvin cell assemblies have been deployed in the past: (i) using gravitational drainage \cite{vanderNet2007} or osmotic effects \cite{Hohler2008} to change the liquid fraction, starting from FCC (face-centered cubic) Kepler foams to obtain BCC (body-centered cubic) Kelvin foams, or (ii) taking advantage of confinement and surface effects: Rosa \textit{et al} \cite{Rosa2002} reorganized layers of bubbles into a Kelvin structure by changing the distance between a liquid interface and a glass plate that confines the bubbles. Meagher \textit{et al} \cite{Meagher2013} used pyramidal containers to modify structures by changing the angle of the pyramid. Confining bubbles into tubes \cite{Saadatfar2008} is also an efficient way to observe a few Kelvin cells, and geometrical constraints imposed by solid walls impose foam structures on a couple of layers against the wall \cite{Skrypnik2023}.

\begin{figure}[h!]
\centering
\includegraphics[width=86mm]{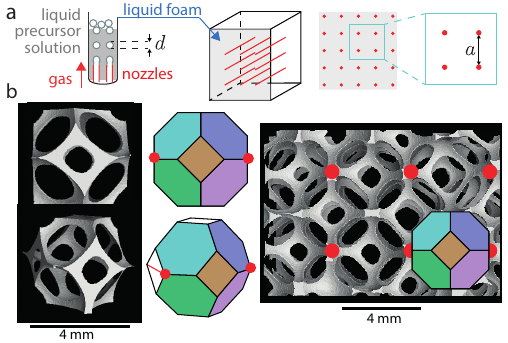}
\caption{a) Experimental setup: gas (air or nitrogen) is blown into a foaming liquid (alginate or polyurethane) through nozzles to form a monodisperse liquid foam, poured into a square cross-section container (50 x 50 x 80 mm) comprising an array of nylon fibers. $d$ is the average bubble diameter and $a$ the fiber spacing. The foams are solidified to analyze their structure. b) Example of long-range ordered structure of Kelvin cells in a BCC configuration, obtained in an alginate foam in a \textit{square array} with spacing close to the bubble diameter (fibers are highlighted in red).}
\label{fig:intro}
\end{figure}

Here we introduce a new strategy with a lighter confinement, considering liquid foams entering an array of rigid fibers (Fig. \ref{fig:intro}a). We show that depending on the ratio of the bubble diameter $d$ to the fiber spacing $a$, architectures made of bubbles constrained by a fiber network can create long-range order and even include Kelvin cells arrangements, as shown in Figure \ref{fig:intro}b.

In two dimensions, the use of pinning centers and of temporal evolution of foams has been used to create well defined 2D structures \cite{Krichevsky1992, Huang2017} including applications in general nanofabrication techniques \cite{Bae2019} and optical materials \cite{Li2017}. In 3D, the work of Whyte \textit{et al} \cite{Whyte2017} has highlighted modifications of structures in simple arrangements of soap films, that are specifically interesting to understand what happens locally. Analysis on the ordering of foams on large scales thanks to fibers, is, to our knowledge, a completely novel route. 

\begin{figure}[b]
\centering
\includegraphics[width=86mm]{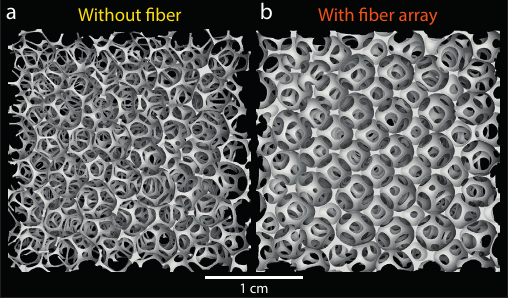}
\caption{a) X-ray tomography scan of an alginate foam without fiber array with bubble diameter $d = 4.3 \pm 0.3$ mm, leading to a disordered structure. b) X-ray tomography scan of an alginate foam in a square fiber array, with bubble diameter $d = 3.7 \pm 0.2$ mm and fiber spacing $a = 4.0$ mm.}
\label{fig:Fig2}
\end{figure}

\begin{figure*}[t]
\centering
\includegraphics[width=172mm]{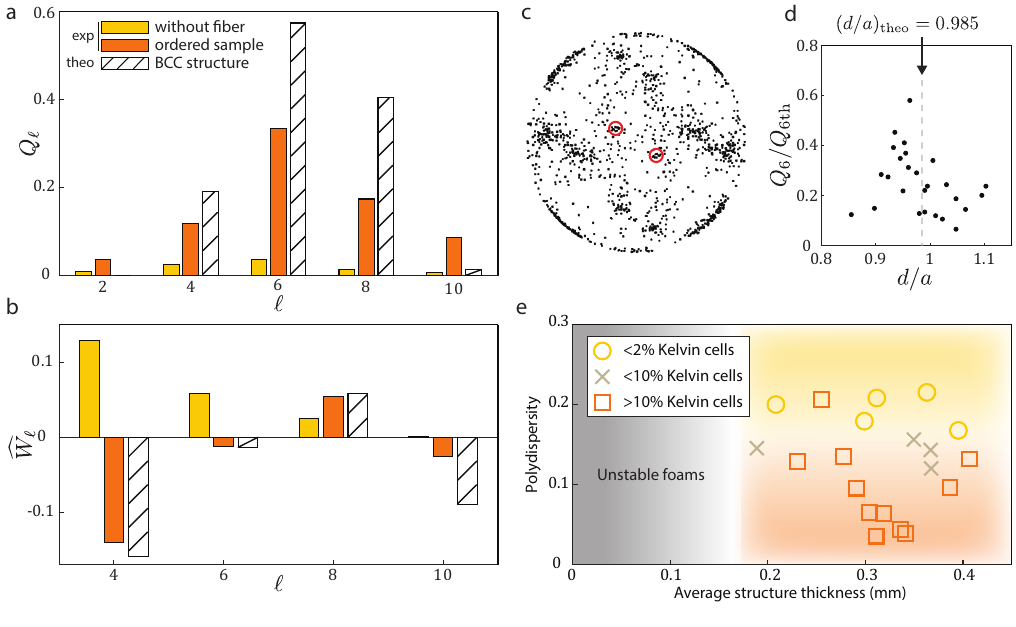}
\caption{a) Coefficients $Q_{\ell}$ extracted in yellow for a sample with no fibers ($d = 4.3 \pm 0.3$ mm), and in orange for the structured zone in a sample with a square fiber array ($d = 3.7 \pm 0.2$ mm, only the skeleton corresponding to the bubbles is kept, struts corresponding to the fibers are removed). Those experimental data are compared to the theoretical values for a BCC arrangement of perfect Kelvin cells (in white, hatched). b) Coefficients $\widehat{W}_{\ell}$ calculated for the same set of systems. Notice that $\widehat{W}_{2}$ is not defined in this case, as the theoretical value of $Q_{2}$ is zero for the structures studied in this work. c) Unit sphere showing the bond orientational density for the ordered sample analyzed in a) and b), highlighting in red the points corresponding to struts actually being fibers. d) Evolution of the obtention of Kelvin cells as a function of the ratio of bubble size to fiber spacing $d/a$, using the ratio $Q_{6}/Q_{6\textrm{th}}$ of the value of $Q_{6}$ obtained experimentally for a cubic central zone of 15 mm side in the sample, divided by the theoretical value $Q_{6\textrm{th}}$ for a perfect Kelvin cell arrangement. $d$ corresponds to the average bubble size over the whole sample. e) Phase diagram highlighting the zones in which clusters of Kelvin cells are obtained, when varying the structure thickness and the polydispersity of bubble sizes. The gray zone on the left corresponds to a zone where stability of foams is not sufficient to allow X-ray tomography scans.}
\label{fig:Steinhardt_Kelvin}
\end{figure*}

\begin{figure}[h!]
\centering
\includegraphics[width=86mm]{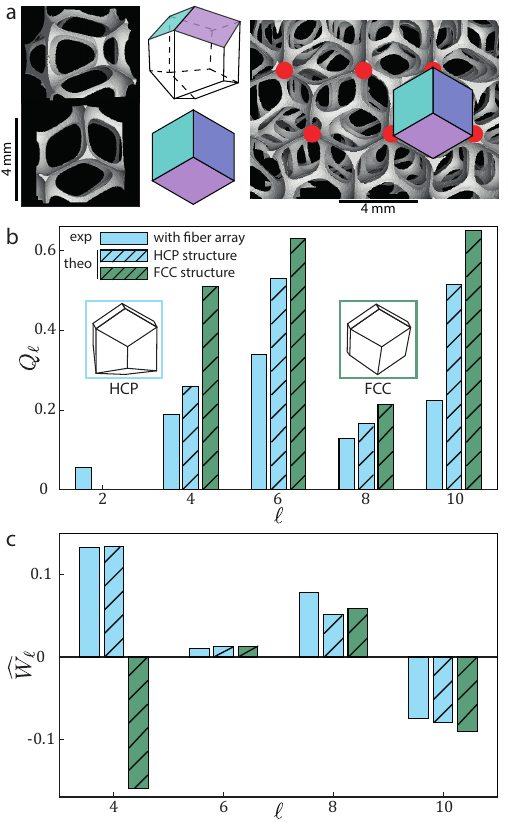}
\caption{a) Obtention of a HCP crystalline structure with a hexagonal fiber array: X-ray tomographic scan of the structure, together with single cell views and the corresponding schematics. b) Coefficients $Q_{\ell}$ calculated in light blue for the structured zone of a sample in a hexagonal fiber array, with fiber spacing $a=4$ mm and bubble diameter $d = 4.2 \pm 0.7$ mm, compared to the theoretical values obtained for a perfect HCP arrangement (hatched light blue) and for a perfect FCC arrangement (hatched green). Insets: schematics of the HCP and FCC cells. c) Coefficients $\widehat{W}_{\ell}$ calculated for the same set of systems.}
\label{fig:Steinhardt_hcp}
\end{figure}

Air or nitrogen bubbles of constant size  are introduced in an initially liquid mixture, and the obtained liquid foam is poured into a container in which a nylon fiber array has been woven (Fig. \ref{fig:intro}a), before undergoing solidification. The air fraction is tuned by controlling drainage, and the bubble diameter $d$ is extracted from the tomography scans as an equivalent diameter (diameter of a sphere of volume equal to the bubble volume). Observations have been made with two polymeric materials to highlight the generality of the approach: polyurethane, shown in the Supplemental Material (Section B), and alginate, which has the advantage of being non toxic and is used as a model system for all results shown below (Fig. \ref{fig:Fig2}, \ref{fig:Steinhardt_Kelvin} and \ref{fig:Steinhardt_hcp}). The general principles of foam preparation are detailed in the absence of fibers in our previously published work (for both alginate \cite{Jouanlanne2024} and polyurethane \cite{Jouanlanne2024b}). Details specific to the current work are described in the Supplemental Material (Section A). The use of such foams that are initially liquid and solidify within the fiber network allows us to (i) analyze the structures via X-ray tomography with no significant drainage as it would be the case for liquid foams \cite{Cantat2013}, and (ii) demonstrate that we can create solid hierachical polymeric materials.

Both visual observations and X-ray tomography scans show that bubbles in a \textit{square} array of fibers can arrange into large clusters of Kelvin cells (truncated octahedrons) in a BCC configuration. Fig. \ref{fig:Fig2} illustrates the difference between foams produced without (a) and with (b) a fiber array of square arrangement with appropriate $d/a$ ratio. The basic physical hypothesis for this ordering are first that the need to collectively minimize the interfacial energy leads the assembly of bubbles to place itself in a rather ordered way so that all fibers are embedded in the liquid part, with a loose pinning of the bubbles to the fibers, similar to the one observed in \cite{Whyte2017}. This "preordering" then biases the path of the foam  within the space of different configurations and allows it to find the Kelvin structure among the wealth of similar-in-energy but highly disordered other metastable equilibriums.

After the capture of the 3D structures of the solidified bubble assemblies via X-ray tomography, we perform a numerical extraction of the strut network with a in-house built matlab code, inspired by the work of Montminy et al \cite{Montminy2001, Montminy2004}. Once the strut network is extracted, we quantify the orientational order in the sample. To do that, we adapted the methodology developed by Steinhardt, Nelson and Ronchetti to study the correlation of rotational order in 3D liquid systems \cite{Steinhardt1981,Steinhardt1983}, which was an extension of the work performed on the prediction of the hexatic phase of 2D liquids \cite{Kosterlitz1973,Nelson1979}. The Steinhardt parameters have been only occasionally used in the context of foams \cite{Meagher2015} or cellular materials or assemblies of capsules \cite{Jose2015}.
This method consists in looking at the global orientational density function of strut orientation over the unit radius 3D sphere (see Fig. \ref{fig:Steinhardt_Kelvin}c for an example): 
\begin{align}
  \rho(\bm{{n}})=\frac{1}{2N_b}\sum_{\langle i,j\rangle}[\delta(\bm{{n}}-\bm{\hat{r}_{ij}})+\delta(\bm{{n}}-\bm{\hat{r}_{ji}})]
\end{align}
where $\bm{\hat{r}_{ij}}=\bm{r_{ij}}/r_{ij}$ is the directional unit vector corresponding to the strut linking the vertex $i$ to its neighbour $j$. This function is defined for any "trial" unit vector $\bm{n}$ and each configuration of the fluid. The sum is over the $N_b$ pairs of nearest-neighbour vertices $(i,j)$ within the foam (notice that this is different from the Steinhardt parameters for fluids, which are defined for directional unit vectors linking nearest neighbour {\sl particles}). The presence of two delta functions for each pair $\langle i,j\rangle$ reflects that  the bonds are not physically oriented, a symmetry $i\leftrightarrow j$ must be preserved.

\medskip

The function $\rho(\bm{n})$ is then represented by its expansion in terms of the spherical harmonics $Y_{\ell}^m(\bm{n})=Y_\ell^m(\theta,\phi)$, assuming an external coordinate system  at the center of the sphere
\begin{align}
  \rho(\bm{n})&=\sum_{\ell=0}^\infty\sum_{m=-\ell}^\ell \rho_{\ell m}{Y_\ell^m}(\theta,\phi),\\
  \rho_{\ell,m}&=\frac{1+(-1)^\ell}{2N_b}\sum_{\langle i,j\rangle}(Y_{\ell}^m)^\star(\theta_{ij},\phi_{ij},)
\end{align}
where $(\theta,\phi)$ are the polar and azimuthal angles of the spherical coordinates of $\bm{n}$, and $z^\star$ is the complex conjugate of $z$. 
Notice that $\rho_{\ell,m}=0$ if $\ell$ is odd, as a consequence of the $\bm{n}\leftrightarrow -\bm{n}$ symmetry.

If no bond orientational order is present in the strut network of the foam, the bond vectors $\bm{\hat{r}_{ij}}$ are expected to uniformly populate the unit sphere, up to finite size fluctuations $\propto 1/\sqrt{N_b}$. For such an isotropic case, all coefficients but the first $\rho_{00}=\frac{1}{2\sqrt{\pi}}$ vanish. Conversely, if an orientational order develops, this isotropy is broken, and the set of mode amplitudes $\rho_{\ell,m}$ will account for this order. As they depend on the choice of the coordinate system,  Steinhardt {\it et al.} proposed to use only combinations of modes which are naturally rotationally invariant (consequently, they will also give identical signatures for any two transited structures having broken the rotational isotropy in different "directions"). The first series of these "Steinhardt parameters" is defined by the discrete function
\begin{align}
  Q_\ell&=\left[\frac{4\pi}{2\ell+1}\sum_{m=-\ell}^\ell |\rho_{\ell,m}|^2\right]^{1/2},
\end{align}
considered only for even integers $\ell$, since $Q_{\rm odd}=0$. Notice also that $Q_\ell=\delta_{\ell,0}$ for an isotropic foam. Another Steinhardt indicator is defined by the cubic combination 
\begin{align}W_{\ell}=\sum_{m_1+m_2+m_3=0}\begin{pmatrix}\ell &\ell&\ell\\m_1&m_2&m_3\end{pmatrix}\rho_{\ell, m_1}\rho_{\ell, m_2}\rho_{\ell, m_3},
\end{align}
(where the matrix is the Wigner $3j$ symbol), or its normalized version $\widehat{W}_\ell=(W_\ell/Q_\ell^3)(4\pi/(2\ell+1))^{3/2}$ which requires that $Q_\ell\neq0$. The combined inspection of $Q_\ell$ and $W_\ell$ gives a rather univocal signature of different periodic layouts of cells and is therefore a useful tool for characterizing partially or totally ordered foams.

Fig. \ref{fig:Steinhardt_Kelvin} presents the results of this analysis on samples with square fiber arrays of spacing $a = 4.0$ mm, compared to samples with no fibers. The coefficients $Q_{\ell}$ (Fig. \ref{fig:Steinhardt_Kelvin}a) and $\widehat{W}_{\ell}$ (Fig. \ref{fig:Steinhardt_Kelvin}b) are computed theoretically for a perfect Kelvin cell arrangement in white (hatched), and compared to a sample with no fibers in yellow, and an ordered part of a sample with a fiber array in orange. Note that the skeleton parts that correspond to the fibers in the foam have been removed, to keep only the bubble network for the analysis. For this foam-fiber assembly, the fibers are not engulfed in struts (Fig. \ref{fig:intro}b). Results with and without removing those parts of the skeleton corresponding to the fibers are presented in the Supplemental Material (Fig. S3). 

We observe that the values of $Q_{\ell}$ for $\ell \neq 0$ with no fibers are close to zero, which is the signature of an isotropic sample (no specific orientational order), whereas the samples exhibiting visually a Kelvin arrangement show a signal much closer to the theoretical values. The values of $Q_{\ell}$ are rather sensitive to noise, however, we still recover a shape of signal for the $Q_{\ell}$ and $\widehat{W}_{\ell}$ that is a clear signature of Kelvin cell arrangements in a BCC configuration. Fig. \ref{fig:Steinhardt_Kelvin}c plots the density function over the unit sphere for the sample with a square fiber array used for Fig. \ref{fig:Steinhardt_Kelvin}a and b, with the fibers highlighted in red.

We then investigate the range of parameters in which such Kelvin arrangements can be observed. The first characterization, shown in Fig. \ref{fig:Steinhardt_Kelvin}d, consist in varying the ratio of the bubble size over the fiber spacing $d/a$. Note that $d$ corresponds to the diameter of a sphere of same volume as the bubble. We observe that Kelvin cells are mostly observed near an optimal value of $d/a$, which is close to the calculated optimum $d/a=0.985$ (see Supplemental Material for the calculation and tomographic images). Note that the values of $d$ correspond to an average over the whole foam, and that the polydispersity is comprised between 3 and 21 \% depending on the sample.

We set the bubble size near this optimum by choosing a nozzle inner diameter of $250$  $\mu$m and construct a phase diagram (Fig. \ref{fig:Steinhardt_Kelvin}e) showing the role of the structure thickness (defined as the average thickness of elementary constituents of the material network (mainly struts), measured by the X-ray tomography analysis software VG Studio MAX \cite{Tomo2022}) and of the sample polydispersity. Details are provided in Table S1 of the Supplemental Material. Several zones can be observed in Fig. \ref{fig:Steinhardt_Kelvin}e: (i) thin structures (highlighted in gray) are not stable enough for X-ray tomography, (ii) polydisperse foams (yellow) are disordered, and (iii) only sufficiently monodisperse foams (orange) exhibit a significant amount of Kelvin cells. Note that we manage to obtain Kelvin cells even for thick structures: the highest material fraction measured by the X-ray tomography software is 7.2 \%, corresponding to a liquid content for which BCC Kelvin foams without fiber are higher in energy than FCC foams \cite{Hohler2008}.

When changing the way fibers are arranged from a square array to an \textit{hexagonal} array, we observe a HCP (hexagonal close packing) array of trapezorhombic dodecahedron bubbles - again for a narrow range of bubble diameter to spacing ratio $d/a$ (see Supplemental Material for results with different values of $d$, Fig. S5). The ordering obtained with the optimal $d/a$ is highlighted on tomographic images in Fig. \ref{fig:Steinhardt_hcp}a. The Steinhardt's coefficients $Q_{\ell}$ and $\widehat{W}_{\ell}$ are shown for the ordered zone of a sample in a hexagonal fiber array in blue, and compared with the theoretical values for a perfect HCP arrangement of trapezorhombic dodecahedron bubbles in hatched blue and a perfect FCC arrangement of rhombic dodecahedron bubbles in hatched green. The discrimination between the two structural arrangements is clear: the obtained structure is a HCP arrangement. It is interesting to note the crucial role played here by $\widehat{W}_{4}$ which alone allows a clear-cut disambiguation between structures. Although the energies of HCP and FCC are equal in the limit of low liquid content, and slightly lower for the HCP structure for wet foams \cite{Whyte2015, Hutzler2020}, FCC seems to be generally prefered in experimental systems \cite{Heitkam2012, Meagher2015}. We show here a strategy to induce preferentially a HCP ordering.

In conclusion, this work demonstrates that taking advantage of the confinement induced by a fiber array can lead to crystalline foam structures usually challenging to observe experimentally. The final structure (Kelvin foam (BCC), or HCP foam) is determined mainly by the fiber pattern and by ratio of the bubble diameter to fiber spacing. The use of polymeric materials able to solidify paves the road towards novel fabrication techniques of highly ordered solid foams: The present desktop scale experiments could be suitable for miniaturisation, with potential applications in acoustic, elastic or photonic metamaterials \cite{Huang2020,Spadoni2014,Klatt2019}, and multiple fiber configurations could be investigated to extend the zoology of foam structures reached through similar processes.

\paragraph{Acknowledgements}
We are grateful to Imene Ben-Djemaa and Léandro Jacomine for their help with the alginate foams formulation and to Wiebke Drenckhan, François Schosseler, Simon Cox, Christian Gauthier and Reinhard Höhler for fruitful discussions. This work of the Interdisciplinary Institute HiFunMat, as part of the ITI 2021-2028 program of the University of Strasbourg, CNRS and Inserm, was supported by IdEx Unistra (ANR-10-IDEX-0002) and SFRI (STRAT’US project, ANR-20-SFRI-0012) under the framework of the French Investments for the Future Program. We also acknowledge funding from the IdEx Unistra framework (A. Hourlier-Fargette) and from the ANR (FOAMINT project, ANR-23-CE06-0014-01).\\

\bibliographystyle{apsrev4-2}
\bibliography{references}
\end{document}